\documentclass[a4paper,12pt,onecolumn,epsfig,floats,amssymb,amsmath]{article}
\usepackage{amsmath}
\usepackage[dvips]{graphicx}

\begin{document}
\title{Stretching An Anisotropic DNA}
\author{B. Eslami-Mossallam and M.R. Ejtehadi \footnote{ejtehadi@sharif.edu}\\ Department of Physics, Sharif University of Technology, \\P.O. Box 11155-9161, Tehran, Iran.}

\bibliographystyle{unsrt}
\maketitle

\begin{abstract}
We present a perturbation theory to find the response of an
anisotropic DNA to the external tension. It is shown that the
anisotropy has a nonzero but small contribution to the
force-extension curve of the DNA. Thus an anisotropic DNA behaves
like an isotropic one with an effective bending constant equal to
the harmonic average of its soft and hard bending constants.
\end{abstract}

\section{Introduction}
One of the most successful theories to describe the physical
behavior of a long DNA molecule is the elastic rod model
\cite{Marko2}. In this theory, the DNA is modeled as a continuous
rod with intrinsic twist (to account for the helical structure of
DNA) which changes its conformation in response to external
forces or torques. The response of the DNA to an external stress
is then mainly determined by three parameters: two principal
bending constants and a twist constant. It is usually assumed that
bending energy is isotropic.

Recent stretching experiments \cite{Smith, Wang, Baumann, Wenner}
allow us to study mechanical response of a single DNA molecule.
Marko and Siggia \cite{Marko} reproduced the measured
force-extension curve of DNA using the isotropic elastic rod model
with an isotropic bending constant of about $50\,\mathrm{nm}$.

Because of DNA special structure, its bending energy is expected
to be anisotropic. The existence of anisotropy in the bending of
DNA has been previously reported by simulation studies as well
\cite{mergell, Lankas}. However, the exact values of the bending
constants in the \emph{easy} and \emph{hard} directions (denoted
here by $A_1$ and $A_2$, respectively) are still unknown.
Recently, Olson \emph{et al.} have stated that the ratio of the
hard bending constant to the easy bending constant is in the
range of $1$ to $5$ \cite{olson2}.

Since the isotropic elastic rod model can explain the observed
force-extension curve in DNA stretching experiments, one may
expect that the response of an anisotropic DNA to the external
tension is similar to an isotropic DNA with an \emph{effective
bending constant}. For a free DNA the effective bending constant
is given by \cite{Maddock}
\begin{equation}
\label{AMaddock}
\frac{1}{A_{\,eff}}=\frac{1}{2}(\frac{1}{A_1}+\frac{1}{A_2})\,.
\end{equation}
We emphasize that the effective bending constant, in fact, depends
on the external constrains applied to DNA. In case of a stretched
DNA, the effective bending constant has been calculated by Nelson
and Moroz \cite{Nelson} only at the large force limit. In this
paper, we present a perturbation theory which allows us to
calculate the force-extension curve of an anisotropic DNA, and
find the effective bending constant.
\section{The Model}
\subsection{The Elastic Rod Model}
\begin{figure}[thb]
\begin{center}
\includegraphics[scale=1]{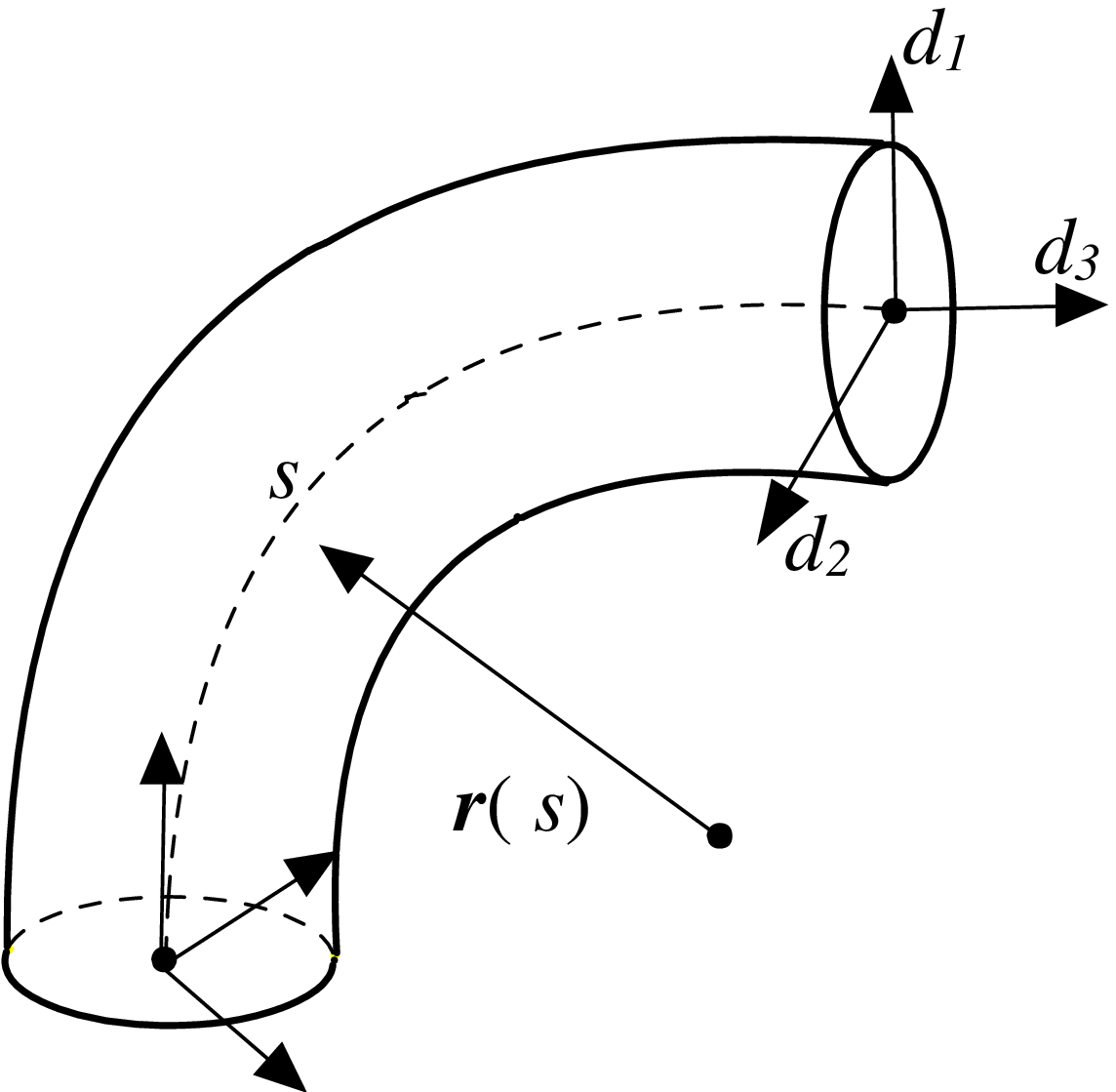}
\end{center} \caption{Parameterization of the elastic rod. The
local frame $\{\hat{d_{1}}, \hat{d_{2}}, \hat{d_{3}}\}$ is
attached to the rod. \label{fig:1}}
\end{figure}

In the elastic rod model the DNA is represented by a continuous
inextensible rod. The curve which passes through the rod center
determines the configuration of the rod in three dimensional
space. This curve is denoted by $\vec{r}$, and is parameterized
by the arc length parameter $s$ (see Figure \ref{fig:1}). In
addition, a local coordinate system with axes $\{\hat{d_{1}},
\hat{d_{2}}, \hat{d_{3}}\}$ is attached to each point of the rod.
$\hat{d_{3}}(s)$ is tangent to the curve $\vec{r}$ at each point
\begin{equation}
\label{d3} \hat{d_{3}}(s)=\frac{\mathrm{d}\vec{r}}{\mathrm{d}s}\,.
\end{equation}
$\hat{d_{1}}(s)$ and $\hat{d_{2}}(s)$ lie in the plane of cross
section of the DNA, and are chosen to be in the easy and hard
directions of bending, respectively.

The orientation of the local coordinate
system with respect to the laboratory coordinate system can be
determined by an Euler rotation defined by
\begin{displaymath}
R(\alpha,\,\beta,\,\gamma)=R_z(\gamma)\,R_y(\beta)\,R_z(\alpha)\,.
\end{displaymath}
$\alpha$, $\beta$, and $\gamma$ are Euler angles. The axes
$\{\hat{d_{1}}, \hat{d_{2}}, \hat{d_{3}}\}$ can then be related to
laboratory coordinate system,
$\{\hat{x},\,\hat{y},\,\hat{z}\}$, via equations
\begin{eqnarray}
\label{d}
\hat{d_{1}}=R^{\,-1}(\alpha,\,\beta,\,\gamma)\,\hat{x}\,,\nonumber\\
\hat{d_{2}}=R^{\,-1}(\alpha,\,\beta,\,\gamma)\,\hat{y}\,,\\
\hat{d_{3}}=R^{\,-1}(\alpha,\,\beta,\,\gamma)\,\hat{z}\,.\,\nonumber
\end{eqnarray}
Thus, if the Euler angles are known as a function of the arc
length parameter $s$, the configuration of the rod will be
uniquely determined.

From classical mechanics we know that
\begin{equation}
\label{ddot} \dot{\hat{d_i}}=\vec{\Omega}\times
\hat{d}_i\quad\quad\quad i=1,2,3\,.
\end{equation}
where the dot denotes the derivative with respect to $s$, and
$\Omega$ is called the \emph{spatial angular velocity}. The
components of $\vec{\Omega}$ in the local coordinate system are
denoted by $\kappa _{1}$, $\kappa _{2}$, and $\omega $
\begin{equation}
\label{Omega} \vec{\Omega}=\kappa _1\;\hat{d_1}+\kappa
_2\;\hat{d_2}+\omega \;\hat{d_3}\,.
\end{equation}
These components can be expressed in terms of Euler angles and
their derivatives with respect to $s$ \cite{Yamakawa}
\begin{eqnarray}
\label{Omega2} \kappa_1=\dot{\beta}\sin \gamma-\dot{\alpha}\sin
\beta \cos \gamma\,\,,\nonumber\\
\kappa_2=\dot{\beta}\cos \gamma+\dot{\alpha}\sin
\beta \sin \gamma\,\,,\\
\omega=\dot{\gamma}+\dot{\alpha}\cos
\beta\,.\qquad\qquad\quad\nonumber
\end{eqnarray}

The elastic rod model introduces the elastic energy as a quadratic
function of $\vec{\Omega}$ components \cite{foot1}
\begin{eqnarray}
\label{Eel} E_{el}=\frac{1}{2}k_BT\int
ds\,\Big[A_1\,\kappa_1^2+A_2\,\kappa_2^2+
C\,(\omega-\omega_0)^2\Big]\,
\end{eqnarray}
where $C$ is the twist constant, and $\omega_0$ is the intrinsic
twist of DNA. the integral is over the entire length of the DNA.
The first two terms in equation (\ref{Eel}) correspond to the
bending of DNA in the easy and hard directions, respectively.
$A_1$ and $A_2$ are the corresponding bending constants $(A_1\leq
A_2)$. Note that the bending energy is isotropic for $A_1=A_2$.
The third term indicates the energy needed for twisting the DNA
about its central axis. 

%
\subsection{Partition Function of a Stretched DNA}
In this section we present a standard method
\cite{Marko,Nelson,Yamakawa,Doi} to calculate the statistical
distribution function of the Euler angles, and to relate this
distribution function to the partition function of a stretched
DNA.
We consider here the case of pure stretching, that is,
stretching with zero applied torque. This situation is realized
in many experiments \cite{Smith, Wang, Baumann, Wenner}.
Also we assume all elastic modulus sequence independent and consider them  to be constant.

Suppose that a DNA molecule is stretched by force $\vec{f}$ along
$\mathbf{\mathrm{\hat{z}}}$ axis.
Following \cite{Marko, Nelson} we neglect self-avoidance effects.
Thus the DNA in our model behaves like a phantom chain. we
also neglect the electrostatic interactions, which are small if
the salt concentration is high enough \cite{Smith, Baumann,
Wenner} . Then,
total energy of DNA can be written as the sum of elastic energy
and the potential energy associated with the tensile force
\begin{equation}
\label{Etot1} E_{tot}=E_{el}-fz\,,
\end{equation}
where $z$ is the end-to-end extension of DNA in the direction of
the external force and is given by
\begin{equation}
\label{z1} z=\int \hat{d_{3}}\cdot
\mathbf{\mathrm{\hat{z}}}\,ds=\int \cos\beta\,ds\,.
\end{equation}
Using equations (\ref{Eel}) and (\ref{z1}), one can write
\begin{equation}
E_{tot}=\int e(s)\,ds\,,
\end{equation}
where $e(s)$  is the energy per unit length of DNA and is given by
\begin{eqnarray}
\label{e}
e(s)=k_BT\Big[\frac{1}{2}\,A_1\,\kappa_1^2+\frac{1}{2}\,A_2\,\kappa_2^2+
\frac{1}{2}\,C\,(\omega-\omega_0)^2 -\tilde{f}\,\cos\beta\Big]\,,
\end{eqnarray}
where $\tilde{f}=\frac{f}{k_BT}$.

It is evident from equations (\ref{Omega2}) and (\ref{e}) that the
DNA total energy depends only on the Euler angles and their
derivatives. This allows us to define a distribution function for
Euler angles. For simplicity, we indicate the three Euler angles
by the vector $\Theta=(\alpha,\,\beta,\,\gamma)$. In order to
obtain the distribution function of $\Theta$, we first define the
\emph{unnormalized} Green function
$\mathbf{G}(\Theta,\,s\,|\,\Theta_0,\,0)$ as follows
\cite{Yamakawa}
\begin{eqnarray}
\label{Green} \mathbf{G}(\Theta_f,\,s\,|\,\Theta_0,\,0)=
\int_{\Theta(0)=\Theta_0}^{\Theta(s)=\Theta_f}
\!\mathcal{D}[\Theta]\exp\!\bigg[\!\!-\frac{1}{k_BT}
\int_{0}^{s}e(s')\,ds'\bigg].
\end{eqnarray}
The path integral in (\ref{Green}) is over all paths between
$\Theta_0$ and $\Theta_f$. We define $\epsilon=\frac{s}{N+1}\,\,$,
$\,s_n=n\,\epsilon$, and $\Theta_n=\Theta(s_n)$. Then the path
integral can be written as
\begin{eqnarray}
\label{path} \int_{\Theta(0)=\Theta_0}^{\Theta(s)=\Theta_f}
\!\mathcal{D}[\Theta]=
\!\!\!\!\!\!\!\!\lim_{\begin{subarray}{l}\\
\ \ \ \ \epsilon\longrightarrow 0\\\\
\ \ \ \ N\longrightarrow\infty\\\\
\ \ \ (N+1)\,\epsilon=s
\end{subarray}}\!\!\!\!\!\Bigg[
\mathcal{N}(\epsilon)\!\!\int d\Theta_1\int d\Theta_2\,\cdots\int
d\Theta_N\Bigg]\,,
\end{eqnarray}
where $d\Theta_n=\sin\beta_n\,d\alpha_n\,\,d\beta_n\,\,d\gamma_n$
\ and
\begin{displaymath}
\mathcal{N}(\epsilon)=[\frac{A_1\,A_2\,C}{(2\pi\epsilon)^{\,3}}]^{\frac{N}{2}}\,.
\end{displaymath}
We call $\mathbf{G}(\Theta,\,s\,|\,\Theta_0,\,0)$ an
unnormalized Green function since the condition
$\int\mathbf{G}(\Theta,\,s\,|\,\Theta_0,\,0)\,d\Theta=1$ is not
satisfied for $f\neq 0$. The unnormalized Green function
 is in fact proportional
to the distribution function of $\Theta$ at point $s$ for $\Theta(0)=\Theta_0$.\\
The above Green function satisfies a Schrodinger-like equation
\cite{Yamakawa}
\begin{equation}
\label{sch1} \left[\frac{\partial}{\partial
s}+H\right]\!\mathbf{G}(\Theta,\,s\,|\,\Theta_0,\,0)=\delta(s)\,\delta(\Theta-\Theta_0)\,,
\end{equation}
where the \emph{Hamiltonian} $H$ is given by
\begin{equation}
\label{H1}
H=\frac{J_1^{\,2}}{2A_1}+\frac{J_2^{\,2}}{2A_2}+\frac{J_3^{\,2}}{2C}+i\,\omega_0\,J_3-\tilde{f}\cos\beta\,,
\end{equation}
with
\begin{eqnarray}
\label{J}
J_1=-i\Big[-\frac{\cos\gamma}{\sin\beta}\,\frac{\partial}{\partial\alpha}+
\sin\gamma\,\frac{\partial}{\partial\beta}+
\cot\beta\,\cos\gamma\,\frac{\partial}{\partial\gamma}\Big]\,,
\nonumber\\
J_2=-i\Big[\quad\frac{\sin\gamma}{\sin\beta}\,\frac{\partial}{\partial\alpha}+
\cos\gamma\,\frac{\partial}{\partial\beta}-
\cot\beta\,\sin\gamma\,\frac{\partial}{\partial\gamma}\Big]\,,\,
\\
J_3=-i\left[\frac{\partial}{\partial\gamma}\right].\qquad\qquad\qquad\qquad\qquad\qquad\qquad\
\quad\,\,\nonumber
\end{eqnarray}
$J_1$, $J_2$, and $J_3$ are analogous to the angular momentum
components of a quantum mechanical top with respect to a
coordinate system attached to it. These angular momentum
components satisfy the commutation relation \cite{Landau}
\begin{equation}
\label{com} [J_i,J_j]=-i\,\epsilon_{ijk}\,J_k\,.
\end{equation}
Note that the term $i\,\omega_0\,J_3$ makes the Hamiltonian
non-Hermitian. In fact the Hamiltonian commutes with the time
reversal operator and belongs to a class of Hamiltonians which
are called pseudo-Hermitian \cite{Mostafazadeh}.

The operators $J_1$ and $J_2$ can also be written in terms of
\emph{ladder operators} $J_{\pm}$
\begin{eqnarray}
\label{J+-} J_1=\frac{1}{2}(J_++J_-)\quad\nonumber\\
J_2=\frac{1}{2i}(J_+-J_-)\,.\,
\end{eqnarray}
Substituting $J_1$ and $J_2$ in equation (\ref{H1}) and using
commutation relation (\ref{com}), we obtain
\begin{eqnarray}
\label{H2} H=
\frac{1}{2A}J^{\,2}+(\frac{1}{2C}-\frac{1}{2A})J_3^{\,2}+\frac{1}{4}\frac{\lambda}{A}(J_+^{\,2}+J_-^{\,2})
+i\,\omega_0\,J_3-\tilde{f}\cos\beta\,,
\end{eqnarray}
here $A$ is the \emph{harmonic average} of $A_1$ and $A_2$
\begin{equation}
\label{A} \frac{1}{A}=\frac{1}{2}(\frac{1}{A_1}+\frac{1}{A_2})\,,
\end{equation}
and
\begin{equation}
\label{lambda}\lambda =\frac{A_2-A_1}{A_1+A_2}\,.
\end{equation}
$\lambda$ is a dimensionless parameter characterizing the anisotropy and varies between zero and one.

We denote the distribution function of $\Theta $ at the point $s$
by $\Psi(\Theta,s)$. From the definition of Green function it is
obvious that $\Psi(\Theta,\,s)$ can be related to $\Psi(\Theta
,\,0)$ via equation
\begin{equation}
\label{Geq} \Psi(\Theta,\,s)=\int
G(\Theta,\,s|\Theta_0,\,0)\,\Psi(\Theta_0,\,0)\,d\Theta_0\,.
\end{equation}
Notice that since the Green function is not normalized, $\Psi(\Theta,s)$ is not normalized either, so we refer to it as the \emph{unnormalized
distribution function}. Considering (\ref{Geq}), $\Psi(\Theta,s)$ also satisfies equation (\ref{sch1})
\begin{equation}
\label{sch2} H\,\Psi(\Theta,s)=-\frac{\partial}{\partial
s}\Psi(\Theta,s)\qquad s>0\,.
\end{equation}
Therefore, we can find $\Psi(\Theta,s)$ by solving the above
Schrodinger-like equation.
\\
We now use Dirac notation to present
our results in a more familiar form. Replacing $\Psi(\Theta,s)$ with $\langle\Theta\,|\Psi(s)\rangle $ we can rewrite equation
(\ref{sch2}) as
\begin{equation}
\label{sch3} H\,|\Psi(s)\rangle=-\frac{\partial}{\partial
s}|\Psi(s)\rangle \,.
\end{equation}

Using equations (\ref{Green}), (\ref{path}), and (\ref{Geq}), the
partition function of a stretched DNA can be written as \cite{Doi}
\begin{equation}
\label{Z1} Z=\int\langle \Theta|\Psi(L)\rangle\, d\Theta\,,
\end{equation}
where $L$ is the total length of DNA. Hence, in order to find the
partition function one must solve the Schrodinger-like
differential equation (\ref{sch2}) and integrate the solution
over all $\Theta$ values.

To solve equation (\ref{sch2}), we rewrite the Hamiltonian in the
form
\begin{equation}
\label{H3} H=H_0+\lambda\,V\,,
\end{equation}
where
\begin{eqnarray}
\label{H0}
H_0=\frac{1}{2A}J^{\,2}+(\frac{1}{2C}-\frac{1}{2A})J_3^{\,2}+i\,\omega_0\,J_3-
\tilde{f}\cos\beta\,,
\end{eqnarray}
and
\begin{equation}
\label{V} V=\frac{1}{4\,A}(J_+^2+J_-^2)\,.
\end{equation}
Furthermore, we decompose $H_0$ to its \emph{real} and
\emph{imaginary} parts
\begin{equation}
\label{H0dec} H_0=H_0^R+iH_0^I\,,
\end{equation}
where
\begin{equation}
\label{H0R}
H_0^R=\frac{J^2}{2\,A}+(\frac{1}{2\,C}-\frac{1}{2\,A})\,J_3^2-\tilde{f}\,\cos\beta
\end{equation}
and
\begin{equation}
\label{H0I} H_0^I=\omega _0\,J_3\,.
\end{equation}
$H_0^{\,R}$ is the Hamiltonian of a quantum top. It commutes with
both $J_3$ and $J_z$, where $J_z$ is the third component of the
angular momentum operator in the laboratory coordinate system
\cite{Landau}. Since $J_3$ and $J_z$ also commute with each
other, one can find the simultaneous eigenvectors of these three
operators. We denote these simultaneous eigenvectors by
$|n,\,k,\,m\rangle$ where $k$ and $m$ are integer numbers
referring to the eigenvalues of $J_3$ and $J_z$, respectively.
The quantum number $n$ distinguishes between the eigenvectors
with identical $k$ and $m$ numbers:
\begin{eqnarray}
\label{eig1}
H_0^R\,\,|n,k,m\rangle=\mathcal{E}_{n,\,k,\,m}^{\,R}\,\,|n,k,m\rangle\,,\\
\label{eig2} J_3\,|n,k,m\rangle=k\,|n,k,m\rangle\,,\qquad\quad\\
\label{eig3} J_z\,|n,k,m\rangle=m\,|n,k,m\rangle\,.\qquad\quad
\end{eqnarray}
From equations (\ref{H0}), (\ref{eig1}), and (\ref{eig2}), it can further be seen that the eigenvectors of $H_0^{\,R}$ are also
eigenvectors of $H_0$:
\begin{equation}
\label{eig4}
H_0\,\,|n,k,m\rangle=\mathcal{E}_{n,\,k,\,m}^{\,0}\,\,|n,k,m\rangle\,,
\end{equation}
where
\begin{equation}
\label{E0}
\mathcal{E}_{n,\,k,\,m}^{\,0}=\mathcal{E}_{n,\,k,\,m}^{\,R}+i\,k\,\omega
_0\,.
\end{equation}
Since $H_0^{\,R}$ is Hermitian, its eigenvectors form a complete
orthogonal basis \cite{foot2}. We now expand $|\Psi (s)\rangle$
in terms of $|n,\,k,\,m\rangle$ eigenvalues
\begin{equation}
\label{expansion} |\Psi
(s)\rangle=\sum_{n,k,m}C_{n,\,k,\,m}(s)\,\,\mathrm{e}^{-\mathcal{E}_{n,\,k,\,m}^{\,0}\,s}\,|n,k,m\rangle
\end{equation}
and substitute $|\Psi (s)\rangle $ into equation (\ref{sch3}). Taking the
orthogonality of the eigenvectors into account we obtain
\begin{eqnarray}
\label{difeq1}  \frac{\partial}{\partial s}\,C_{n,\,k,\,m}=
-\lambda\!\!\sum_{n',k',m'}\langle n,k,m|V|n',k',m'\rangle
\mathrm{e}^{-(\,\mathcal{E}_{n',\,k',\,m'}^{\,0}-\mathcal{E}_{n,\,k,\,m}^{\,0})\,s\,}\,C_{n',\,k',\,m'}\,.
\quad
\end{eqnarray}
The ladder operators in $V$ imply that \cite{Landau}
\begin{eqnarray}
\label{Vmat} \langle n,k,m|V|n',k',m'\rangle =
\langle n,k,m|V|n',k+2,m\rangle\,\delta _{m',m}\,\delta_{k',k+2}+\nonumber\\
\langle n,k,m|V|n',k-2,m\rangle\,\delta
_{m',m}\,\delta_{k',k-2}\,.\
\end{eqnarray}
so we have
\begin{eqnarray}
\label{difeq2} \frac{\partial}{\partial s}\,C_{n,\,k,\,m}=
-\lambda\,\sum_{n'}\langle n,k,m|V|n',k+2,m\rangle
\mathrm{e}^{-(\,\mathcal{E}_{n',\,k+2,\,m}^{\,0}-\mathcal{E}_{n,\,k,\,m}^{\,0})\,s\,}\,C_{n',\,k+2,\,m}
\nonumber\\
-\lambda\,\sum_{n'}\langle n,k,m|V|n',k-2,m\rangle
\mathrm{e}^{-(\mathcal{E}_{n',\,k-2,\,m}^{\,0}-\mathcal{E}_{n,\,k,\,m}^{\,0})\,s}\,C_{n',\,k-2,\,m}\,.
\end{eqnarray}

Substituting  $ |\psi (L)\rangle $ from equation
(\ref{expansion}) into equation (\ref{Z1}) we can derive an
expression for the partition function. Since $\int \langle \Theta
|n,\,k,\,m\rangle\,d\Theta$ is non-zero only for $k=m=0$, we
obtain \cite{Landau}
\begin{equation}
\label{Z2}
Z=\sum_{n}I_n\,C_{n,0,0}(L)\,e^{-\mathcal{E}_{n,\,0,\,0}^{\,0}\,L}
\end{equation}
where
\begin{equation}
\label{I}I_n \equiv\int \langle \Theta |n,\,0,\,0\rangle\,d\Theta
\,.
\end{equation}

Thus, to determine the partition function of a stretched DNA, one
needs to find the coefficients $C_{n,0,0}(L)$ by solving the
differential equation (\ref{difeq2}).
\subsection{Perturbation Theory}
In this section, we use perturbation theory to find the expansion
coefficients and the partition function in powers of $\lambda$.
Let's expand $C_{n,k,m}(s)$ in terms of $\lambda$:
\begin{equation}
\label{Cexpand}
C_{n,k,m}(s)=\sum_{p=0}^{\infty}{\lambda}^p\,C^{\,(p)}_{n,k,m}(s)\,.
\end{equation}
As a result, the partition function can be written as
\begin{equation}
\label{Zexpand} Z=\sum_{p=0}^{\infty}\lambda ^p\,Z^{\,(p)}\,,
\end{equation}
where
\begin{equation}
\label{Z(p)}
Z^{\,(p)}=\sum_{n}I_n\,C^{\,(p)}_{n,0,0}(L)\,e^{-\mathcal{E}_{n,\,0,\,0}^{\,0}\,L}\,.
\end{equation}
By inserting $C_{n,k,m}(s)$ from equation (\ref{Cexpand}) into
equation (\ref{difeq2}), one can see that $C^{\,(p)}_{n,0,0}(s)$
satisfies the following differential equations
\begin{equation}
\label{Cdot0} \frac{\partial}{\partial s}\,C^{\,(0)}_{n,\,k,\,m}=0
\end{equation}
for $p=0$, and
\begin{eqnarray}
\label{Cdotp} \frac{\partial}{\partial s}\,C^{\,(p)}_{n,\,k,\,m}=
-\sum_{n'}\langle n,k,m|V|n',k+2,m\rangle
\mathrm{e}^{-(\,\mathcal{E}_{n',\,k+2,\,m}^{\,0}-\mathcal{E}_{n,\,k,\,m}^{\,0})\,s\,}\,C^{\,(p-1)}_{n',\,k+2,\,m}
\nonumber\\
-\sum_{n'}\langle n,k,m|V|n',k-2,m\rangle
\mathrm{e}^{-(\mathcal{E}_{n',\,k-2,\,m}^{\,0}-\mathcal{E}_{n,\,k,\,m}^{\,0})\,s}\,C^{\,(p-1)}_{n',\,k-2,\,m}
\quad\quad
\end{eqnarray}
for $p>0$.\\
The value of $|\Psi (0)\rangle$ is determined by anchoring the
DNA hence independent of $\lambda$. Thus the corresponding
initial conditions are
\begin{eqnarray}
\label{init}
C^{\,(0)}_{n,k,m}(0)=C_{n,k,m}(0)
\ \quad\,\textrm{for}\ p=0 \,\,\nonumber\\
C^{\,(p)}_{n,k,m}(0)=0\quad \textrm{for}\ p>0\,.\qquad\qquad
\end{eqnarray}

It can be seen from equation (\ref{Cdot0}) that
$C^{\,(0)}_{n,\,k,\,m}$ is constant
\begin{equation}
\label{C0} C^{\,(0)}_{n,k,m}=C_{n,k,m}(0)\,.
\end{equation}
Therefore, the partition
function to the zeroth order of $\lambda$ is given by
\begin{equation}
\label{Z(0)} Z^{\,(0)}=\sum
_{n,k}b^{\,(0)}_{n,k}\,\,e^{-\mathcal{E}_{n,\,k,\,0}^{\,0}\,L}\,,
\end{equation}
where
\begin{equation}
\label{b(0)}
b^{\,(0)}_{n,k}=I_n\,C^{\,(0)}_{n,0,0}\,\,\delta_{k,0}\,.
\end{equation}
$Z^{(\,0)}$ is the partition function of an isotropic DNA with
bending constant $A$.\\
The differential equation (\ref{Cdotp}) can be solved by
iteration, and the corrections to $Z^{(\,0)}$ can be found in
powers of $\lambda$. The first order correction is given by
\begin{equation}
\label{Z(1)} Z^{\,(1)}=\sum
_{n,k}b^{\,(1)}_{n,k}\,\,e^{-\mathcal{E}_{n,\,k,\,0}^{\,0}\,L}\,,
\end{equation}
and the second order correction is given by
\begin{equation}
\label{Z(2)} Z^{\,(2)}=\sum
_{n,k}(b^{\,(2)}_{n,k}-U^{\,(2)}_{n}\,C^{\,(0)}_{n,0,0}\,\delta
_{k,0} \,L)\,\,e^{-\mathcal{E}_{n,\,k,\,0}^{\,0}\,L}\,.
\end{equation}
The coefficients $b^{\,(1)}_{n,k}$ and $b^{\,(2)}_{n,k}$ in
equations (\ref{Z(1)}) and (\ref{Z(2)}) are given in appendix
\ref{appB}. They depend on the initial conditions but do not
depend on the length of DNA. The coefficient $U^{\,(2)}_{n}$ is
given by
\begin{eqnarray}
\label{U(2)} U^{\,(2)}_{n}=\sum_{ n_1\notin
G_{n},\,n_2}I_{n_1}\bigg[\frac{\langle
n_1,\,0|V|n_2,\,2\rangle\,\langle n_2,\,
2|V|n,\,0\rangle}{(\mathcal{E}_{n,\,
0}^{\,0}-\mathcal{E}_{n_2,\,2}^{\,0})}+
\nonumber\\
\frac{\langle n_1,\,0|V|n_2,\,-2\rangle\,\langle n_2,\,
-2|V|n,\,0\rangle}{(\mathcal{E}_{n,\,
0}^{\,0}-\mathcal{E}_{n_2,\,-2}^{\,0})}\bigg],
\end{eqnarray}
where for simplicity, we omit the quantum number $m$ keeping in mind that $m=0$.
$G_{n}$ in equation (\ref{U(2)}) refers to all eigenvectors with eigenvalues equal to
$\mathcal{E}_{n,\,0,\,0}^{\,0}$
\begin{equation}
\label{Gn} n_1 \in G_n \Leftrightarrow
\mathcal{E}_{n_1,\,0,\,0}^{\,0}=\mathcal{E}_{n,\,0,\,0}^{\,0}\,.
\end{equation}
Clearly, if the eigenvector $|n,\,0,\,0\rangle $ is not
degenerate, we have $ G_n=\{n\}$.

The coefficients $b^{\,(1)}_{n,k}$ are zero except for $k=0,\pm 2$
(see appendix \ref{appB}). Since the imaginary part $\mathcal{E}_{n,\,k,\,0}^{\,0}$
is $k\,\omega _0$, an oscillatory term with
frequency $2\,\omega _0$ appears in $Z^{(\,1)}$. In fact,
from equation (\ref{Cdotp}) we expect that oscillatory terms with
frequencies $\{2\,\omega _0,\,4\,\omega _0,\,\cdots\,,2p\,\omega
_0\}$ appear in the expression of $Z^{(\,p)}$. The appearance of
oscillatory terms is, in fact, an artifact of coupling between bending
and twisting in an anisotropic DNA \cite{Farshid}. This is the
main difference between the partition functions of an isotropic
and an anisotropic DNA. Although, as we will show in the next
section, this difference is not detectable in experiments, at
least if the DNA is long enough.
\subsection{The Average End to End Extension \label{AverageExtension}}
Using equations (\ref{Etot1}), (\ref{Green}), (\ref{Geq}), and
(\ref{Z1}) the average end-to-end extension of the DNA can be
calculated as \cite{Marko, Nelson}
\begin{equation}
\label{z2} \frac{\langle z\rangle}{L}=\frac{1}{L}\,\frac{\partial
\ln Z }{\partial \tilde{f}}\,.
\end{equation}
Following Marko and Siggia \cite{Marko}, we limit our study to
the long DNA. In this case, because of the presence of
$\exp(-\mathcal{E}_{n,\,k,\,0}^{\,R}\,L) $ factor,  the term which
corresponds to the ground state eigenvalue of $H_0^{\,R}$ is much
greater than other terms in the expansion of the partition
function. Therefore, the partition function can be approximated
only by the ground state term where all other terms can be
neglected. If we denote the difference between the ground state
and the first excited state eigenvalues by
$\Delta\mathcal{E}^{R}$ then the long DNA limit corresponds to the
condition $\Delta\mathcal{E}^{R}L\gg 1$. We will discuss in the
next section that this condition is indeed satisfied in the
stretching experiments.

The operator $H_0^{\,R}$ is the Hamiltonian of a top in a uniform
external field, and its ground state is unique. Thus the ground
state of $H_0^{\,R}$ must be a simultaneous eigenvector of $J_3$
and $J_z$, with eigenvalues $m=k=0$ \cite{Marko}. We denote the
ground state and its eigenvalue by $|0,0,0\rangle $ and
$\mathcal{E}_{0,\,0,\,0}^{\,R}$ respectively. Therefore, at long
DNA limit we obtain
\begin{equation}
\label{Z(0)long} Z^{\,(0)}\simeq
b^{\,(0)}_{0,0}\,\,e^{-\mathcal{E}_{0,\,0,\,0}^{\,0}\,L}\,,
\end{equation}
\begin{equation}
\label{Z(1)long} Z^{\,(1)}\simeq
b^{\,(1)}_{0,0}\,\,e^{-\mathcal{E}_{0,\,0,\,0}^{\,0}\,L}\,,
\end{equation}
and
\begin{equation}
\label{Z(2)long} Z^{\,(2)}\simeq
(b^{\,(2)}_{0,0}-U^{\,(2)}_{0}\,C^{\,(0)}_{0,0,0}
\,L)\,\,e^{-\mathcal{E}_{0,\,0,\,0}^{\,0}\,L}\,.
\end{equation}
Since the ground state is not degenerate, $ G_0=\{0\}$ and one
can write
\begin{equation}
\label{U(2)0} U_0^{\,(2)}=\mathcal{E}_{0,\,0,\,0}^{\,2}\,I_0\,,
\end{equation}
where
\begin{eqnarray}
\label{E(2)}
\mathcal{E}^{\,2}_{0,0,0}=\sum_{n_2}\bigg[\,\frac{\big|\langle
0,\,0,\,0|\,V|n_2,\, 2,\,0\rangle\big|^{\,2}}{(\mathcal{E}_{0,\,
0,\,0}^{\,0}-\mathcal{E}_{n_2,\,2,\,0}^{\,0})} +
\frac{\big|\langle 0,\,0,\,0|\,V|n_2,\,
-2,\,0\rangle\big|^{\,2}}{(\mathcal{E}_{0,\,
0,\,0}^{\,0}-\mathcal{E}_{n_2,\, -2,\,0}^{\,0})}\,\bigg].
\end{eqnarray}
Therefore, $Z^{\,(2)}$ can be written as
\begin{equation}
\label{Z(2)long2} Z^{\,(2)}\simeq
(b^{\,(2)}_{0,0}-b^{\,(0)}_{0,0}\,\mathcal{E}^{\,2}_{0,0,0}
\,L)\,\,e^{-\mathcal{E}_{0,\,0,\,0}^{\,0}\,L}\,.
\end{equation}

We expand $\langle z\rangle $ in powers of $\lambda $,
\begin{equation}
\label{z-expand} \frac{\langle z\rangle}{L}=\frac{\langle
z^{\,(0)}\rangle}{L}+\lambda \frac{\langle z^{\,(1)}\rangle}{L}+
\lambda ^2 \frac{\langle z^{\,(2)}\rangle}{L}+O(\lambda ^3)\,.
\end{equation}
From equation (\ref{z2}), we have
\begin{equation}
\label{z(0)-1} \frac{\langle
z^{\,(0)}\rangle}{L}=\frac{1}{L}\,\frac{\partial \ln Z^{\,(0)}
}{\partial \tilde{f}}\,,
\end{equation}
\begin{eqnarray}
\label{z(1)-1} \frac{\langle
z^{\,(1)}\rangle}{L}=\frac{1}{L}\,\frac{\partial }{\partial
\tilde{f}}\,\frac{\partial\ln Z}{\partial \lambda}\bigg
|_{\lambda=0}=\frac{1}{L}\,\frac{\partial }{\partial
\tilde{f}}\,\frac{Z^{\,(1)}}{Z^{\,(0)}}\,,\!\!\!\!\quad\quad\qquad\,\,
\end{eqnarray}
and
\begin{eqnarray}
\label{z(2)-1} \frac{\langle
z^{\,(2)}\rangle}{L}=\frac{1}{2\,L}\,\frac{\partial }{\partial
\tilde{f}}\,\frac{\partial ^2\ln Z}{\partial \lambda ^2 }\bigg
|_{\lambda=0}=\frac{1}{L}\,\frac{\partial }{\partial
\tilde{f}}\,\bigg[\frac{Z^{\,(2)}}{Z^{\,(0)}}-\frac{1}{2}\Big(\frac{Z^{\,(1)}}{Z^{\,(0)}}\Big)^2\bigg]\,.
\end{eqnarray}

Neglecting terms of order $\frac{1}{L}$ \cite{Marko, Nelson}, we
obtain
\begin{equation}
\label{z(0)-2} \frac{\langle z^{\,(0)}\rangle}{L}\simeq
-\frac{\partial
 \mathcal{E}^{\,0}_{0,0,0}}{\partial \tilde{f}}\,,
\end{equation}
\begin{equation}
\label{z(1)-2} \frac{\langle z^{\,(1)}\rangle}{L}\approx 0\,,\
\qquad\,\,
\end{equation}
\begin{equation}
\label{z(2)-2} \frac{\langle z^{\,(2)}\rangle}{L}\simeq
-\frac{\partial
 \mathcal{E}^{\,2}_{0,0,0}}{\partial \tilde{f}}\,.
\end{equation}

So far we have assumed that DNA is
inextensible. To account for the extensibility of DNA the term
$\frac{\tilde{f}}{B}$ must be added to $\frac{\langle
z\rangle}{L}$ , where $B\,k_BT$ is the stretch modulus of DNA and
is about $500\,\mathrm{k_BT\,nm^{-1}}$ \cite{Marko}. Thus one can
write
\begin{equation}
\label{z(0)-3} \frac{\langle z^{\,(0)}\rangle}{L}\simeq
-\frac{\partial
 \mathcal{E}^{\,0}_{0,0,0}}{\partial
 \tilde{f}}+\frac{\tilde{f}}{B}\,.
\end{equation}
$\langle z^{\,(0)}\rangle $ is the average end-to-end extension
of an isotropic DNA with the bending constant $A$.  Marko and
Siggia have also calculated $\langle z^{\,(0)}\rangle $
\cite{Marko}. Although they used a different Hamiltonian, i.e.
\begin{displaymath}
H_{iso}=\frac{J^2}{2\,A}-\tilde{f}\cos\beta\,,
\end{displaymath}
our results are identical to theirs to the zeroth order. The
reason is that $H_0^{\,R}$ and $H_{iso}$ have the same ground
state eigenvalues.
\section{Results \label{results}}
\begin{figure}[htb]
\begin{center}
\includegraphics[width=1\textwidth]{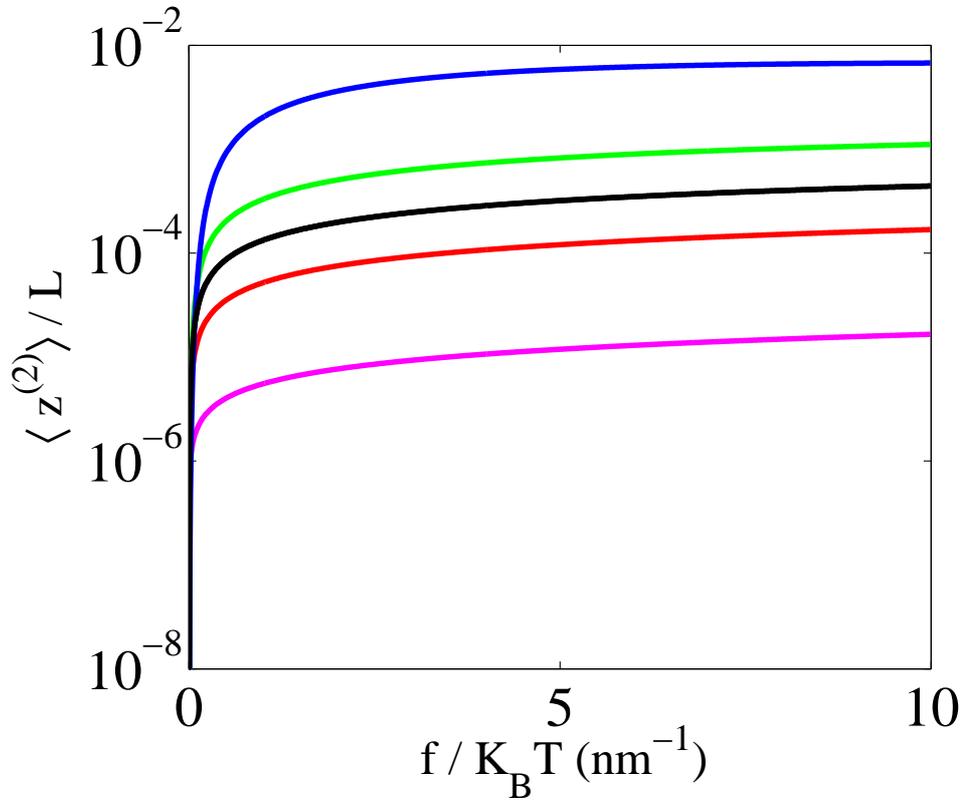}
\end{center}
\caption{(color online) The contribution of the second order
term to the relative extebtion of DNA as a function of
$\tilde{f}$, for $C=100\,\mathrm{nm}$, $\omega
_0=1.8\,\mathrm{nm^{-1}}$. The carves show the result for
different values of $A$. From top to bottom $A=5\,\mathrm{nm}$
(Blue),  $A=25\,\mathrm{nm}$ (green), $A=50\,\mathrm{nm}$
(black),  $A=100\,\mathrm{nm}$ (red), and $A=500\,\mathrm{nm}$
(magenta), respectively.\label{fig:2}}
\end{figure}

Numerical methods are employed (see appendix \ref{appA}) to
calculate the second order correction to the force extension
curve of an isotropic DNA, assuming $A=50\,\mathrm{nm} $,
$C=100\,\mathrm{nm}$ \cite{Nelson}, and $\omega
_0=1.8\,\mathrm{nm^{-1}}$. The result is shown in Figure
\ref{fig:2}. For forces slightly greater than $\tilde{f}\sim
10\,\mathrm{nm^{-1}}$ the DNA undergoes an over-stretching
transition \cite{Smith2}, hence the elastic rod model is not
relevant. We have therefore picked the force range of
$0<\tilde{f}<10\,\mathrm{nm^{-1}}$ to insure validity.

\begin{figure}[htb]
\begin{center}
\includegraphics[width=1\textwidth]{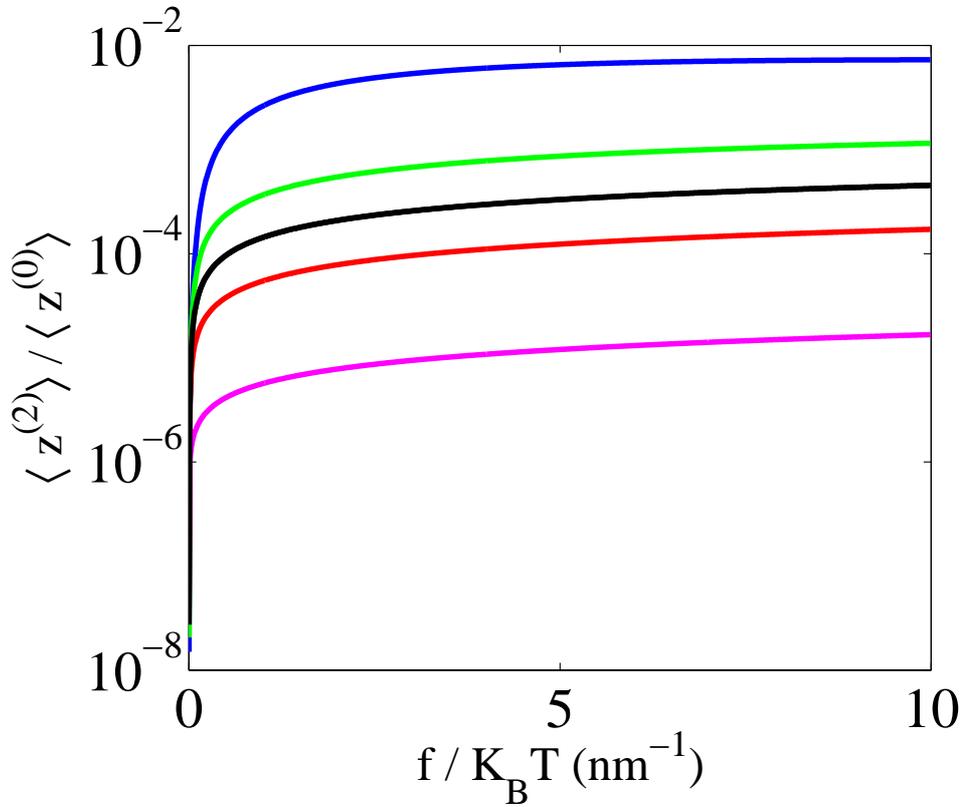}
\end{center}
\caption{(Color online) The ratio $\frac{\langle z^{\,(2)}\rangle}{\langle
z^{\,(0)}\rangle}$ as a function of $\tilde{f}$ for the curves
shown in Figure \ref{fig:2}. From top to bottom
$A=5\,\mathrm{nm}$ (Blue),  $A=25\,\mathrm{nm}$ (green),
$A=50\,\mathrm{nm}$ (black),  $A=100\,\mathrm{nm}$ (red), and
$A=500\,\mathrm{nm}$ (magenta), respectively.
\label{fig:3}}
\end{figure}

It can be seen from Figure \ref{fig:2} that $\langle
z^{\,(2)}\rangle $ is positive. Therefore, to the second order of
$\lambda$, anisotropy increases the average extension of DNA. However,
$\langle z^{\,(2)}\rangle $ is small compared to $\langle
z^{\,(0)}\rangle $. For $A=50\,\mathrm{nm}$, the maximum value of
the ratio $\frac{\langle z^{\,(2)}\rangle}{\langle
z^{\,(0)}\rangle}$ is in the order of $10^{-4}$ (see Figure \ref{fig:3}).\\
To be sure that this result is not limited to the special case of
$A=50\,\mathrm{nm}$, we examine four different values of $A$ in
the range $5\leq A\leq 500\,\mathrm{nm}$ (see Figure
\ref{fig:2}). The ratio  $\frac{\langle z^{\,(2)}\rangle}{\langle
z^{\,(0)}\rangle}$ for these four values of $A$ are plotted in
Figure \ref{fig:3}. It can be seen that $\frac{\langle z^{\,(2)}\rangle}{\langle z^{\,(0)}\rangle}$
does not exceed $10^{-2}$ for $A \geq 5 \mathrm{nm}
$.

As can be seen from Figure \ref{fig:2}, for
$A=50\,\mathrm{nm}$, where the theoretical curve is best fitted
to the experimental data \cite{Marko}, one must measure
$\frac{\langle z\rangle}{L}$ at least with the accuracy $10^{-4}$
to detect $\langle z^{\,(2)}\rangle $ . Since $L\sim
10\,\mathrm{\mu m}$ in experiments \cite{Smith}, minimum accuracy
of $1\,\mathrm{nm}$ is required in measuring $\langle z\rangle $.
However, the accuracy of the experiments is by far less than this
limit \cite{Smith}, therefore $\langle z^{\,(2)}\rangle $ can not
be detected by stretching experiments.

We now show that $\langle z^{\,(3)}\rangle $ is also small. It is
obvious that when $\Psi (\Theta ,0)$ is independent of the Euler
angle $\gamma$, the partition function is invariant under the
transformation $\lambda\rightarrow -\lambda $. This means that
odd powers of $\lambda $ are not present in the expansion of
$\langle z\rangle $, i.e., $\langle z^{\,(2p+1)}\rangle =0$. In
addition, the effect of the initial conditions on the force
extension curve of DNA is suppressed if DNA is long enough. As a
result, one expects  $\frac{\langle z^{\,(2p+1)}\rangle}{L}$ to
be small even when $\Psi (\Theta ,0)$ depends on $\gamma $.  In
other words, odd powers of $\lambda$ have no significant
contribution to the end-to-end DNA extension. Therefore, to the
third order of $\lambda$, the response of an anisotropic DNA to
the external tension is close to an isotropic DNA with the
effective bending constant
\begin{equation}
\label{Aeff}
A_{eff}=A=2\left(\frac{1}{A_1}+\frac{1}{A_2}\right)^{-1}\,.
\end{equation}

\begin{figure}[htb]
\begin{center}
\includegraphics[width=1\textwidth]{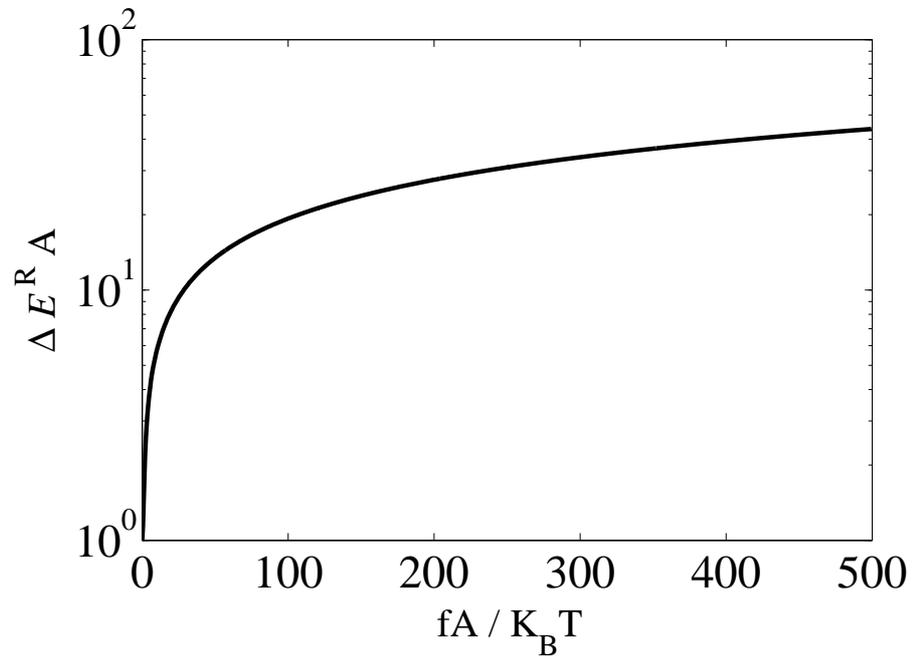}
\end{center}
\caption{$\Delta\mathcal{E}^{R}A$ as a function of $\tilde{f}A$
for $A=50\,\mathrm{nm}$, $C=100\,\mathrm{nm}$, and $\omega
_0=1.8\,\mathrm{nm^{-1}}$.\label{fig:4}}
\end{figure}

To justify our result, we must show that the condition
$\Delta\mathcal{E}^{R}L\gg 1$ which corresponds to the limit of
long DNA, is satisfied in experiments as well. Figure \ref{fig:4} shows
$\Delta\mathcal{E}^{R}A$ as a function of $\tilde{f}A$ for
$A=50\,\mathrm{nm}$. As can be seen, $\Delta\mathcal{E}^{R}A\geq
1$. As a result, the condition $\Delta\mathcal{E}^{R}L\gg 1$ is
equivalent to the condition $L\gg A$, which is well known in
polymer physics. Since $A=50\,\mathrm{nm}$ and $L\sim
10\,\mathrm{\mu m}$, this condition is satisfied in the streching
experiments.

\section{Discussion and Conclusion}
It is well known that when  DNA is free (i.e., no external force
applied), the average energy of an anisotropic DNA is equal to
the average energy of an isotropic DNA with bending constant $A$
\cite{Lankas}. Moreover, Maddocks and Kehrbaum \cite{Maddock} have
proved that in the absence of external forces or torques the
ground state configuration of an anisotropic DNA is similar to the
ground state configuration of an isotropic DNA with bending
constant $A$. However, a stretched DNA is not free. More
importantly, to calculate the average end-to-end extension one
must deal with the \emph{free energy} instead of the average
energy or the ground state energy.

The partition function of a stretched DNA is generally represented as
\begin{equation}
\label{general Z}
Z=\int_{E_{min}}^\infty\mathcal{D}(E)\,\exp(-\frac{E}{k_BT})\,\mathrm{d}E\,,
\end{equation}
where $\mathcal{D}(E)\,\mathrm{d}E$ is the number of possible
configuration with an energy in the range of $E$ and
$E+\mathrm{d}E$, and $E_{min}$ is the ground state energy. For an
stretched DNA, the ground state corresponds to the configuration
in which the DNA is fully stretched. The equilibrium
configuration of the DNA is the configuration that minimize the
free energy, $F=E-k_BT\ln \mathcal{D}(E)$, and therefore is
different from the ground state configuration. Clearly, bending
anisotropy changes $\mathcal{D}(E)$ for \emph{excited
configurations} thus changes the free energy and equilibrium
configuration of the stretched DNA. When no external force is
applied to the DNA, the number of configurations that have the
end-to-end extension $z$ is exactly equal to the number of
configurations with the end-to-end extension $-z$. Consequently
we have $\langle z\rangle=0$ regardless of the degree of
anisotropy, $\lambda$. Thus in the limit of $\tilde{f}A\ll 1$,
anisotropy can barely affect the average end-to-end extension,
and one expects $\langle z^{\,(2)}\rangle$ and in fact all the
higher-order corrections to be small, as can be seen from Figures
\ref{fig:2} and \ref{fig:3}. On the other hand, in the limit of
$\tilde{f}A\gg 1$, the energy of the ground state is much lower
than those of the excited states, and the excited configurations
have a small contribution to the partition function. Therefore,
the effect of anisotropy will be suppressed at large forces, and
$\langle z^{\,(2)}\rangle$ and all the higher-order corrections
vanish as $\tilde{f}A\rightarrow\infty$. This is the reason that
$\frac{\partial \langle z^{\,(2)}\rangle}{\partial f}$ is smaller
at large forces (see Figure \ref{fig:2}).

Nelson and Moroz \cite{Nelson} have applied an approximate method to
obtain an analytical expression for $\langle z\rangle $ at the
limit of large forces to the second order of $\lambda $. They
found
\begin{displaymath}
A_{eff}=\bar{A}(1-2(\frac{\hat{A}}{\stackrel{}{\bar{A}}})^2)=A-\lambda
^2\, \bar{A}\,,
\end{displaymath}
with $\bar{A}=\frac{1}{2}(A_1+A_2)$, and
$\hat{A}=\frac{1}{2}(A_2-A_1)$. This result is different
from equation (\ref{Aeff}). However, we rederived their calculations and obtained the same result as in equation (\ref{Aeff})
\begin{displaymath}
A_{eff}=\bar{A}(1-(\frac{\hat{A}}{\stackrel{}{\bar{A}}})^2)=A
\end{displaymath}
Thus we believe that they just made an error in their calculations. It can be shown that the result of these calculations is
in fact exact (see appendix \ref{app3}). Therefore, at the high
force limit, an anisotropic DNA behaves like an isotropic DNA
with the bending constant $A$.

\section{Acknowledgement}
We thank N. Hamedani Radja for insightful conversations and his valuable comments on analytical calculations. We do also thank H. Amirkhani for her comments on the draft manuscript and Behzad Eslami-Mossallam for the figures illustrations. MRE thanks the Center of Excellence in Complex Systems and Condensed Matter (CSCM) and the Institute for studies in Theoretical Physics and Mathematics for their partial supports.

\appendix

\section{Expansion Coefficients for $Z^{\,(1)}$ and $Z^{\,(2)}$ \label{appB}}
Here we present the expressions for $b^{\,(1)}_{n,k}$ and
$b^{\,(2)}_{n,k}$. Let's use the following abbreviations
\begin{equation}
\label{abbreviation1} V_{n,\,k,\,n',\,k'}\equiv \langle
n,\,k,\,0|\,V| n',\,k',\,0\rangle\,,
\end{equation}
and
\begin{equation}
\label{abbreviation2} \Delta \mathcal{E}_{n,\,k,\,n',\,k'}^{\,0}=
\mathcal{E}_{n,\,k,\,0}^{\,0}-\mathcal{E}_{n',\,k',\,0}^{\,0}\,.
\end{equation}
The $b^{\,(1)}_{n,k}$ coefficients are
\begin{eqnarray}
\label{b(1)-1} b^{\,(1)}_{n,\pm
2}=\sum_{n_1}I_{n_1}\,\frac{V_{n_1,\,0,\,n,\,\pm 2}}{\Delta
\mathcal{E}_{n,\,\pm 2,\,n_1,\,0}^{\,0}}\,\,C_{n_1,\pm 2,0
}^{(0)},\qquad\qquad\qquad\quad\quad\quad\\
\label{b(1)-2} b^{\,(1)}_{n,0}= I_n\,\sum_{n_1}\,\Big[\frac{
V_{n,\,0,\,n_1,\,2}}{\Delta\mathcal{E}_{n,\,0,\,n_1,\,2}^{\,0}}\,\,C_{n_1,2,0
}^{(0)}+ \frac{
V_{n,\,0,\,n_1,\,-2}}{\Delta\mathcal{E}_{n,\,0,\,n_1,\,-2}^{\,0}}\,\,C_{n_1,-2,0
}^{(0)}\Big],
\end{eqnarray}
and
\begin{equation}
 \label{b(1)-3}
 b^{\,(1)}_{n,k}=0 \qquad k\neq 0,\,\pm
2\,.\qquad\qquad\qquad\qquad\qquad\qquad\qquad\quad\,\,\,\,
\end{equation}
 The $b^{\,(2)}_{n,k}$ coefficients are
\begin{eqnarray}
\label{b(2)-1} b^{\,(2)}_{n,\pm 4}= \sum_{n_1,n_2}I_{n_1}\Big[
\frac{V_{n_1,\,0,\,n_2,\,\pm 2}\,\,V_{n_2,\,\pm 2,\,n,\,\pm
4}}{\Delta \mathcal{E}_{n,\,\pm 4,\,n_1,\,0}^{\,0}\,\,\Delta
\mathcal{E}_{n,\,\pm 4,\,n_2,\,\pm 2}^{\,0}}\Big]C_{n,\pm
4,0}^{(0)},
\end{eqnarray}
\begin{eqnarray}
\label{b(2)-2} b^{\,(2)}_{n,\pm 2}= \sum_{n_1,n_2}I_{n_2}\Big[
\frac{V_{n_2,\,0,\,n,\,\pm 2}\,\,V_{ n,\,\pm 2,\,n_1,\,\pm
4}}{\Delta \mathcal{E}_{n,\,\pm 2,\,n_1,\,\pm 4}^{\,0}\,\,\Delta
\mathcal{E}_{n,\,\pm 2,\,n_2,\,0}^{\,0}
}\,\,C_{n_1,\pm 4,0}^{(0)}\nonumber\\
+ \frac{V_{n_2,\,0,\,n,\,\pm 2}\,\, V_{n,\,\pm
2,\,n_1,\,0}}{\Delta \mathcal{E}_{n,\,\pm 2,\,n_1,\,0
}^{\,0}\,\,\Delta \mathcal{E}_{n,\,\pm 2,\,n_2,\,0}^{\,0}
}\,\,C_{n_1,0,0}^{(0)}\Big],
\end{eqnarray}
\begin{eqnarray} \label{b(2)-3}
b^{\,(2)}_{n,0}= \sum_{n_1,\,n_2}\sum_{k=\pm 2}I_n \Big[
\frac{V_{n,\,0,\,n_2,\, k}\,\,V_{n,\, k,\,n_1,\,
2k}}{\Delta\mathcal{E}_{n_1,\,
2k,\,n_2,\,k}^{\,0}\,\,\Delta\mathcal{E}_{n,\,0,\,n_1,\,2k}^{\,0}}
\qquad\quad\nonumber\\
-\frac{V_{n,\,0,\,n_2,\, k}\,\,V_{n,\, k,\,n_1,\,
2k}}{\Delta\mathcal{E}_{n_1,\,
2k,\,n_2,\,k}^{\,0}\,\,\Delta\mathcal{E}_{n,\,0,\,n_2,\,k}^{\,0}}
\Big]C_{n_1,2k,0}^{(0)}\nonumber\\
-\sum_{n_1,\,n_2}\sum_{k=\pm 2}I_{n}\Big[\frac{V_{n,\,0,\,n_2,\,
k}\,\, V_{n_2,\,k,\,n_1,\,0}}{\Delta \mathcal{E}_{n,\,
0,\,n_2,\,k}^{\,0}\,\,\Delta \mathcal{E}_{n_1,\,0,\,n_2,\,k}^{\,0}}\Big]C_{n_1,0,0}^{(0)}\nonumber\\
+\sum_{\begin{subarray}{l} n_1,\,n_2\\
n_1\notin G_{n}\end{subarray}}\sum_{k=\pm 2}I_{n_1}\Big[
\frac{V_{n_1,\,0,\,n_2,\,k}\,\,V_{n_2,\, k,\,n,\,0}}{\Delta
\mathcal{E}_{n,\,
0,\,n_2,\,k}^{\,0}\,\,\Delta\mathcal{E}_{n,\,0,\,n_1,\,0}^{\,0}}\Big]C_{n,0,0}^{(0)}\nonumber\\
-\sum_{\begin{subarray}{l} n_1,\,n_2\\
n_1\notin G_{n}\end{subarray}}\sum_{k=\pm 2}I_n\Big[
\frac{V_{n,\,0,\,n_2,\,k}\,\,V_{n_2,\, k,\,n_1,\,0}}{\Delta
\mathcal{E}_{n_1,\, 0,\,n_2,\,k}^{\,0}\,\,\Delta
\mathcal{E}_{n_1,\,0,\,n,\,0}^{\,0}}\Big]C_{n_1,0,0}^{(0)},
\end{eqnarray}
and
\begin{eqnarray}
\label{b(2)-4} b^{\,(2)}_{n,k}=0 \qquad\qquad k\neq 0,\,\pm
2,\,\pm 4\,.\qquad\qquad\qquad\qquad\,\,
\end{eqnarray}
In equation (\ref{b(2)-3}), $\sum_{k=\pm 2}$ indicates that one
must sum over both $k=2$ and $k=-2$.
\section{Numerical Calculations\label{appA}}
To calculate the eigenvectors and eigenvalues of $H_0^{\,R}$, we
use the eigenvectors of the angular momentum operator as the
basis of the Hilbert space. We denote these eigenvectors by
$|\chi_{\,j,\,k,\,m}\rangle $. From quantum mechanics, one knows
that $|\chi_{\,j,\,k,\,m}\rangle $ satisfies the following
eigenvalue equations \cite{Landau}
\begin{equation}
\label{eigj2}
J^2|\chi_{\,j,\,k,\,m}\rangle=j\,(j+1)|\chi_{\,j,\,k,\,m}\rangle
\quad\!\!\! j\in \mathbf{Z}^+\cup \{0\},
\end{equation}
\begin{equation}
\label{eigj3}
J_3\,|\chi_{\,j,\,k,\,m}\rangle=k\,|\chi_{\,j,\,k,\,m}\rangle
\quad |k|\leq j\,,\qquad\qquad\quad
\end{equation}
\begin{equation}
\label{eigjz}
J_z\,|\chi_{\,j,\,k,\,m}\rangle=m\,|\chi_{\,j,\,k,\,m}\rangle
\quad |m|\leq j\,.\qquad\qquad\quad
\end{equation}
The vector $|n,\,k,\,0\rangle $ can be expanded in terms of
$|\chi_{\,j,\,k,\,0}\rangle $ as
\begin{equation}
\label{eigenvecExpansion} |n,\,k,\,0\rangle =\sum_{j=k}^{\infty}
a_{n,\,j,\,k}\,|\chi_{\,j,\,k,\,0}\rangle \,.
\end{equation}
Then the equation
$H_0^{\,R}\,\,|n,k,0\rangle=\mathcal{E}_{n,\,k,\,0}^{\,R}\,\,|n,k,0\rangle
$ transforms to the matrix equation
\begin{eqnarray}
\label{MATeq} \sum_{j'=k}^{\infty}\Big[\langle
\chi_{\,j,\,k,\,0}|H_0^{\,R}|\chi_{\,j',\,k,\,0}\rangle -
\mathcal{E}_{n,\,k,\,0}^{\,R}\,\delta_{j,\,j'}\Big]a_{n,\,j',\,k}=0\,,
\end{eqnarray}
where $\langle
\chi_{\,j,\,k,\,0}|H_0^{\,R}|\chi_{\,j',\,k,\,0}\rangle $ is given
by \cite{Landau, Arfken}
\begin{eqnarray}
\label{Hmatrix}\langle
\chi_{\,j,\,k,\,0}|H_0^{\,R}|\chi_{\,j',\,k,\,0}\rangle=
\bigg[\frac{1}{2\,A}\,j(j+1)+
\frac{1}{2}(\frac{1}{C}-\frac{1}{A})\,k^2\bigg]\,\delta_{j,\,j'}-
\qquad\qquad\quad\nonumber\\
\tilde{f}\,\bigg[\frac{\sqrt{(j'-k)(j'+k)}\,\,\delta_{j,\,j'-1}}{\sqrt{(2j+1)(2j'+1)}}+
\frac{\sqrt{(j-k)(j+k)}\,\,\delta_{j',\,j-1}}{\sqrt{(2j+1)(2j'+1)}}\bigg].
\end{eqnarray}
Similarly, $\langle n,\,k,\,0|V|n',\,k',\,0\rangle$ can be written
as
\begin{eqnarray}
\label{VMateq} \langle n,\,k,\,0|V|n',\,k',\,0\rangle =
\sum_{j=k}^\infty \sum_{j'=k'}^\infty
a_{n,\,j,\,k}^{\,\star}\,a_{n',\,j',\,k'}\,\langle
\chi_{\,j,\,k,\,0}|V|\chi_{\,j',\,k',\,0}\rangle\,,
\end{eqnarray}
where $\langle \chi_{\,j,\,k,\,0}|V|\chi_{\,j',\,k',\,0}\rangle $
is given by \cite{Landau}
\begin{eqnarray}
\label{Vmatrix} \langle
\chi_{\,j,\,k,\,0}|V|\chi_{\,j',\,k',\,0}\rangle =
\qquad\qquad\qquad\qquad\qquad\qquad\qquad\qquad\qquad\qquad\nonumber\\
\frac{1}{4\,A}\,\delta_{j,\,j'}\,\delta_{k,\,k'+2}
\bigg[\sqrt{(j+k)\,(j-k+1)(j+k-1)\,(j-k+2)}
\bigg]+\nonumber\\
\frac{1}{4\,A}\,\delta_{j,\,j'}\,\delta_{k,\,k'-2}
\bigg[\sqrt{(j+k')\,(j-k'+1)(j+k'-1)\,(j-k'+2)}\bigg].
\end{eqnarray}

The dimension of the matrix
$\langle\chi_{\,j,\,k,\,0}|H_0^{\,R}|\chi_{\,j',\,k,\,0}\rangle $
is infinite. Thus, to solve the eigenvalue equation (\ref{MATeq})
numerically, we choose a cutoff for $j$. We find that the
calculated values for $\mathcal{E}_{0,\,0,\,0}^{\,0}$ and
$\mathcal{E}_{0,\,0,\,0}^{\,2}$ converge very rapidly. A choice
of $j_{max}=120$ is sufficient to calculate $\langle
z^{\,(0)}\rangle $ and $\langle z^{\,(2)}\rangle $ with a
relative accuracy of $10^{-8}$ (taking into account the error due
to numerical differentiating).

\section{Average End To End Extension of the DNA at large force Limit\label{app3}}
If the external tension is adequately large, the DNA remains relatively
straight. Thus, $\hat{d}_{3}$ lies approximately in the
$\mathbf{\mathrm{\hat{z}}}$ direction and
$\hat{d}_{1}$ and $\hat{d}_2$ will be confined, as a result, in the $xy$ plane. In this
case, the components of the spatial angular velocity can be
written in this form
\begin{eqnarray}
\label{app3-omegacomp} \kappa_1=-\hat{d}_{3x}\sin\phi -
\hat{d}_{3\,y}\cos\phi\,\,,\nonumber\\
\kappa_2=\hat{d}_{3\,x}\cos\phi +
\hat{d}_{3\,y}\sin\phi\,,\quad\\
\omega =\frac{\mathrm{d}\phi}{\mathrm{d}s}\,.
\qquad\qquad\qquad\quad\quad\,\,\,\,\nonumber
\end{eqnarray}
Where $\phi (s)$ is the twist angle of DNA, and $\hat{d}_{3\,x}$
and $\hat{d}_{3\,y}$ are the components of $\hat{d}_{3}$ in the
$\mathbf{\mathrm{\hat{x}}}$ and
$\mathbf{\mathrm{\hat{y}}}$ directions, respectively:
\begin{displaymath}
\hat{d}_{3}=\hat{d}_{3\,x}\mathbf{\mathrm{\hat{x}}}+
\hat{d}_{3\,y}\mathbf{\mathrm{\hat{y}}}+\hat{d}_{3\,z}\mathbf{\mathrm{\hat{z}}}\,.
\end{displaymath}
Further, since $\hat{d}_{3\,x}$ and $\hat{d}_{3\,y}$ are both small, we can write
\begin{equation}
\label{app3-d3z}
\hat{d}_{3\,z}\approx1-\frac{1}{2}(\hat{d}_{3\,x}^{\,2}+\hat{d}_{3\,y}^{\,2})\,.
\end{equation}
Defining $\bar{A}=\frac{1}{2}(A_1+A_2)$ and
$\hat{A}=\frac{1}{2}(A_2-A_1)$, the energy of DNA can be
written as
\begin{equation}
\label{app3-Etot} E=E_0+E_1+E_{\mathrm{twist}}-fL\,,
\end{equation}
where
\begin{eqnarray}
\label{app3-E0} \frac{E_{0}}{k_{B}T}= \frac{1}{2}\int_{0}^L
\left[\
\bar{A}(\dot{\hat{d}}_{3\,x}^{\,2}+\dot{\hat{d}}_{3\,y}^{\,2})+\tilde{f}(\hat{d}_{3\,x}^{\,2}+\hat{d}_{3\,y}^{\,2})\
\right]ds,
\end{eqnarray}
\begin{eqnarray}
\label{app3-E1-1} \frac{E_1}{k_{B}T}= \frac{\hat{A}}{2}\int_{0}^L
\left[\cos(2\phi)
(\dot{\hat{d}}_{3\,x}^{\,2}-\dot{\hat{d}}_{3\,y}^{\,2})\right]ds\nonumber\\
+ \frac{\hat{A}}{2}\int_{0}^L
\left[2\sin(2\phi)\,\dot{\hat{d}}_{3\,x}\,\dot{\hat{d}}_{3\,y}\right]ds\,,
\end{eqnarray}
and
\begin{equation}
\label{app3-Etw}\frac{E_{\mathrm{twist}}}{k_BT}=\frac{C}{2}\int_0^L
(\omega - \omega_0)^2\,ds\,.
\end{equation}

On the basis of the \emph{ergodic} principle, one expects that the
relation
\begin{displaymath}
\frac{1}{s}\int_0^s \omega (s')ds'=\langle \omega\rangle =\omega_0
\end{displaymath}
holds for large $s$ \cite{foot3}. Thus, for a long DNA we can
employ the approximation \cite{foot4}
\begin{equation}
\label{app3-phi} \phi (s)\simeq \omega _0 s\,,
\end{equation}
and substitute $\phi (s)$ into equation (\ref{app3-E1-1}) to get
\begin{eqnarray}
\label{app3-E1-2} \frac{E_1}{k_{B}T}= \frac{\hat{A}}{2}\int_{0}^L
\left[\cos(2\omega _0 s)
(\dot{\hat{d}}_{3\,x}^{\,2}-\dot{\hat{d}}_{3\,y}^{\,2})\right]ds\nonumber\\
+ \frac{\hat{A}}{2}\int_{0}^L \left[2\sin(2\omega _0
s)\,\dot{\hat{d}}_{3\,x}\,\dot{\hat{d}}_{3\,y}\right]ds\,.
\end{eqnarray}

To calculate the partition function, we express the total energy
in terms of Fourier components of $\hat{d}_{3\,x}$ and
$\hat{d}_{3\,y}$. The Fourier transform of
$\hat{d}_{3\,x}+i\,\hat{d}_{3\,y}$ is given by
\begin{equation}
\label{app3-Ft}
\hat{d}_{3\,x}+i\,\hat{d}_{3\,y}=\sum_{j=-\infty}^\infty a_j\,
\exp(i\, q_j\, s)\,,
\end{equation}
where $q_j=\frac{2\,j\,\pi}{L}$. Using the properties of Fourier
transformation, we obtain \cite{Nelson}
\begin{equation}
\label{app3-FE0} E_0=\frac{L}{2}\,k_B
T\!\!\!\sum_{j=-\infty}^\infty (\bar{A}q_j^2+\tilde{f})\,|\,a_j|^2
\end{equation}
and
\begin{equation}
\label{app3-FE1} E_1=-\frac{\hat{A}}{4}L\,k_B
T\!\!\!\sum_{j=-\infty}^\infty
q_j\,q_{j_0-j}(a_j\,a_{j_0-j}\,+\,c.c.),
\end{equation}
where $q_{j_0}$ is the closest \emph{wave number} to $2\omega_0$
\begin{displaymath}
2\omega_0\simeq \frac{2\,\pi\,j_0}{L}\,.
\end{displaymath}
We denote the real and imaginary parts of $a_j$ as $R_j$, and $I_j$
,respectively. Then the total energy of the DNA can be written in
the form
\begin{equation}
\label{app3-FE} E=-fL+E^R+E^I+E_{\mathrm{twist}}\,,
\end{equation}
with
\begin{eqnarray}
\label{app3-ER}\frac{E^R}{k_BT}=\frac{L}{2}\sum_{j=-\infty}^\infty
(\bar{A}q_j^2+\tilde{f})\,R_j^2
-\frac{\hat{A}L}{2}\sum_{j=-\infty}^{\infty}q_j\,q_{j_0-j}\,R_j
R_{j_0-j}\,,
\end{eqnarray}
and
\begin{eqnarray}
\label{app3-EI}\frac{E^I}{k_BT}=\frac{L}{2}\sum_{j=-\infty}^\infty
(\bar{A}q_j^2+\tilde{f})\,I_j^2
+\frac{\hat{A}L}{2}\sum_{j=-\infty}^{\infty}q_j\,q_{j_0-j}\,I_j
I_{j_0-j}\,,
\end{eqnarray}
Therefore, the partition function is given by
\begin{equation}
\label{app3-Z} Z=Z_R \,Z_I \,Z_{\mathrm{twist}}\,,
\end{equation}
where
\begin{equation}
\label{app3-ZR}
Z_R=\exp(\frac{\tilde{f}L}{2})\int_{-\infty}^{\infty}\prod_{j=-\infty}^\infty
dR_j\,\exp(-\frac{E^R}{k_BT}),
\end{equation}
\begin{equation}
\label{app3-ZI}
Z_I=\exp(\frac{\tilde{f}L}{2})\int_{-\infty}^{\infty}\prod_{j=-\infty}^\infty
dI_j\,\exp(-\frac{E^I}{k_BT}),
\end{equation}
and
\begin{equation}
\label{app3-Ztw} Z_{\mathrm{twist}}=\int
\mathcal{D}[\omega]\,\exp(-\frac{E_{\mathrm{twist}}}{k_BT})\,.
\end{equation}
The integral in equation (\ref{app3-Ztw}) is taken over all
possible paths of $\omega (s)$.

The average end-to-end extension of DNA can be calculated from
equation (\ref{z2}). Since $Z_{\mathrm{twist}}$ does not depend
on $f$, one can write
\begin{equation}
\label{app3-z1} \langle z\rangle =\frac{\partial}{\partial
\tilde{f}}(\ln Z_R + \ln Z_I)\,.
\end{equation}
It is clear from equations (\ref{app3-ER}) and (\ref{app3-EI})
that one needs to calculate only $Z_I$. $Z_R$ can be calculated
simply by replacing $\hat{A}$ with $-\hat{A}$ in the expression
obtained for $Z_I$.

From equation (\ref{app3-EI}) we have
\begin{equation}
\label{app3-EIdec} E^I=\left\{\begin{array}{ll} \sum_{j\geq
\frac{j_0+1}{2}}^\infty E^I_j & \textrm{$j_0$ is odd.}\\
\\
\sum_{j> \frac{j_0}{2}}^\infty E^I_j + E^I_0 & \textrm{$j_0$ is
even.}
\end{array}\right.
\end{equation}
where
\begin{eqnarray}
\label{app3-EIj}\frac{E^I_j}{k_BT}=\frac{L}{2}\,\big[
(\bar{A}q_j^2+\tilde{f})\,I_j^2+(\bar{A}q_{j_0-j}^2+\tilde{f})\,I_{j_0-j}^2
+\,2\hat{A}\,L\,q_j\,q_{j_0-j}\,I_j I_{j_0-j}\big],
\end{eqnarray}
and
\begin{equation}
\label{app3-EI0}\frac{E^I_0}{k_BT}= \frac{L}{2}\,\big[
(\bar{A}+\hat{A})q_{\frac{j_0}{2}}^2+\tilde{f}\big]\,I_{\frac{j_0}{2}}^2\,.
\end{equation}
For simplicity we assume that $j_0$ is odd. It can easily be
shown that the final result does not change when $j_0$ is even.
Since the variables $I_j$ and $I_{j_0-j}$ only appear in
$E^I_j$, substitution of $E^I$ in equation (\ref{app3-ZI}) yields
\begin{eqnarray}
\label{app3-ZI-2} Z_I=\exp(\frac{\tilde{f}L}{2})\prod_{j=j\geq
\frac{j_0+1}{2}}^\infty\Big[\int_{-\infty}^{\infty}
dI_j\,dI_{j_0-j}\,\exp(-\frac{E^I_j}{k_BT})\Big].
\end{eqnarray}
The integrals in equation (\ref{app3-ZI-2}) can be calculated
using the formula
\begin{eqnarray}
\int_{-\infty}^{\infty}dx\,dy\, \exp[-a\,x^2-b\,y^2+2\,c\,x\,y]=
\frac{\pi}{\sqrt{a\,b-c^2}}\,.
\end{eqnarray}
Then we obtain
\begin{eqnarray}
\label{app3-logZ} \ln Z_{I}=\ln Z_R=\frac{1}{2}\ln Z_0-
\frac{1}{4}\sum_{j=-\infty}^\infty \ln
\left[1-\hat{A}^2\,F(q_j)\,F(q_{j_0-j})\right],
\end{eqnarray}

where
\begin{eqnarray}
\label{app3-Z0} Z_0=\exp(\tilde{f}L)
\int_{-\infty}^{\infty}\Big[\prod_{j=-\infty}^\infty
dI_j\,dR_{j}\Big]\,\exp(-\frac{E_0}{k_BT})
\end{eqnarray}
is the partition function of an isotropic DNA with bending
constant $\bar{A}$ and
\begin{equation}
\label{F} F(q)\equiv \frac{q^2}{\bar{A}q^2+\tilde{f}}\,.
\end{equation}
Using equations (\ref{app3-z1}) and (\ref{app3-logZ}),the average
end-to-end extension of DNA is given by
\begin{eqnarray}
\label{app3-z2}  \frac{\langle z\rangle}{L}=\frac{{\langle
z\rangle}_0}{L}-\frac{1}{2L}\sum_{j=-\infty}^\infty
[G(q_j)+G(q_{j_0-j})],
\end{eqnarray}
where $\langle z\rangle _0$ is the average end-to-end extension
of an isotropic DNA with the bending constant $\bar{A}$
\cite{Marko, Nelson},
\begin{equation}
\label{app3-z0} \frac{{\langle z\rangle}_0}{L}= \frac{1}{L}
\frac{\partial\ln Z_0}{\partial
\tilde{f}}=1-\frac{1}{2\,\sqrt{\tilde{f}\bar{A}}},
\end{equation}
and
\begin{eqnarray}
\label{app3-G} G(q)=
\frac{\hat{A}^2\,F(q)\,F(q-2\omega_0)}{(\bar{A}q^2+\tilde{f})(1-\hat{A}^2\,F(q)\,F(q-2\omega_0))}\,.
\end{eqnarray}
The sum in equation (\ref{app3-z2}) can be transformed into an
integral as follows
\begin{eqnarray}
\label{app3-z3}  \frac{\langle z\rangle}{L}=\frac{{\langle
z\rangle}_0}{L}-\frac{1}{4\pi}\int_{-\infty}^\infty
[G(q)+G(2\omega_0-q)]\,dq
\qquad\nonumber\\
=\frac{{\langle
z\rangle}_0}{L}-\frac{1}{2\pi}\int_{-\infty}^\infty
G(q)\,dq\,.\qquad\qquad\qquad\qquad
\end{eqnarray}
Changing the  integration variable $q$ to
$x=\sqrt{\frac{\bar{A}}{\stackrel{}{\tilde{f}}}}\,q$, and defining
$x_0=2\sqrt{\frac{\bar{A}}{\stackrel{}{\tilde{f}}}}\,\omega_0$ and
$\lambda =\frac{\hat{A}}{\stackrel{}{\bar{A}}}$, one can write
\begin{eqnarray}
\label{app3-intG} \int_{-\infty}^\infty
G(q)dq=\frac{\lambda^2}{\sqrt{\tilde{f}\bar{A}}}
\int_{-\infty}^\infty
\frac{U(x)\,U(x-x_0)}{(x^2+1)(1-\lambda^2\,U(x)\,U(x-x_0))}\,dx\,,
\end{eqnarray}
with
\begin{equation}
\label{app3-U} U(x)=\frac{x^2}{1+x^2}.
\end{equation}
From equations (\ref{app3-z0}), (\ref{app3-z3}) and
(\ref{app3-intG}) we obtain
\begin{equation}
\label{app3-z4} \frac{\langle z\rangle}{L}=
1-\frac{1}{2\,\sqrt{\tilde{f}\bar{A}}}\,(1+g(\lambda)),
\end{equation}
where $g(\lambda)$ is given by
\begin{eqnarray}
\label{app3-g} g(\lambda)=\frac{\lambda^2}{\pi}
\int_{-\infty}^\infty
\frac{U(x)\,U(x-x_0)}{(x^2+1)(1-\lambda^2\,U(x)\,U(x-x_0))}\,dx,\,
\end{eqnarray}
Since $x_0\gg 1$ in the range of experimental data, we employ
Nelson and Moroz approximation \cite{Nelson}
\begin{equation}
\label{app3-approximateU} U(x-x_0)\simeq 1,
\end{equation}
to calculate the integral in equation (\ref{app3-g}). We find
\begin{equation}
\label{app3-g2} g(\lambda)=\frac{1}{\sqrt{1-\lambda^2}}-1.
\end{equation}
Thus we obtain
\begin{equation}
\label{app3-z5} \frac{\langle
z\rangle}{L}=1-\frac{1}{2\sqrt{\tilde{f}\bar{A}}}
\left(1-(\frac{\hat{A}}{\stackrel{}{\bar{A}}})^2\right)^{-\frac{1}{2}}.
\end{equation}
Comparing equation (\ref{app3-z5}) with equation (\ref{app3-z0}),
one can see that the effective bending constant is given by
\begin{equation}
\label{app3-Aeff}
A_{eff}=\bar{A}\,\left(1-(\frac{\hat{A}}{\stackrel{}{\bar{A}}})^2\right)=
2\left(\frac{1}{A_1}+\frac{1}{A_2}\right)^{-1}.
\end{equation}
This is the same result that we have obtained in section
\ref{results}.
\bibliography{biblio}

\begin{thebibliography}{10}
\bibitem{foot1}
In this and the next two section, we assume that the DNA is inextensible. The
  extensibility of DNA is taken into account in section \ref{AverageExtension}.
  Also we have neglected the twist-bend and stretch-twist couplings, 
  well as the effect of sequence on DNA elasticity. 
\bibitem{foot2}
Since the Hamiltonian $H$ is not Hermitian, there is no guarantee that its
  eigenvectors form a complete basis. So we choose the eigenvectors of
  $H_0^{\,R}$ as the basis of the Hilbert space.
\bibitem{foot3}
The probability that the twist of the DNA is equal to $\omega $ in the range
  $s_0<s<s_0+ \Delta s_0$, is proportional to the factor $\exp[-C(\omega -
  \omega_0)^2\Delta s]$. One can see that, if $(C\,\Delta s)^{-1}\ll \omega
  _0^2$, the probability that $\omega$ differs from $\omega_0$ is negligible.
  Therefore the DNA can not maintain a constant twist $\omega \neq \omega_0$,
  at distances which are much greater than $ \Delta
  s_{max}=(C\,\omega_0^{2})^{-1}$. As a result, for $s\gg \Delta s_{max}$, the
  integral $\int_0^s \omega (s')ds'$ contain many different values of $\omega $
  which are distributed about $\omega_0$, and the ergodic principle holds.
  Since $\Delta s_{max}\sim 10^{-3}\,\mathrm{nm}$ and $L\sim
  10\,\mu\mathrm{m}$, the equation (\ref{app3-phi}) is a good approximation for
  $\phi $.
\bibitem{foot4}
Since the initial conditions can not affect the force-extension relation for a
  long DNA, we assume $\phi (0)=0$ for simplicity.
\end{thebibliography}
\newpage
\listoffigures
\newpage

\newpage

\newpage

\newpage

\end{document}